\documentclass[aps,prl,twocolumn,groupedaddress,showpacs]{revtex4}
\usepackage{bm}
\usepackage{graphicx,amssymb}
\usepackage{amsmath}

\newcommand{\beq}{\begin{equation}}
\newcommand{\eeq}{\end{equation}}
\newcommand{\beqa}{\begin{eqnarray}}
\newcommand{\eeqa}{\end{eqnarray}}

\begin{document}

\title{Classical Model for Jellium}
\author{Sandipan Dutta and James Dufty}
\affiliation{Department of Physics, University of Florida\\
Gainesville, FL 32611}
\date{\today }
\pacs{ }

\begin{abstract}
A simple, practical model for computing the equilibrium
thermodynamics and structure of jellium by classical strong coupling
methods is proposed. An effective pair potential and coupling
constant are introduced, incorporating the ideal gas, low density,
and weak coupling quantum limits. The resulting parameter free,
analytic model is illustrated by the calculation of the pair
correlation function over a wide range of temperatures and densities
via strong coupling classical liquid state theory. The results
compare favorably with the first finite temperature restricted path
integral Monte Carlo simulations reported recently.
\end{abstract}

\date[Date text]{date}
\maketitle

The limitations of many body theories for strongly coupled quantum
systems at finite temperatures have led to attempts to adapt
corresponding methods known to be effective for classical systems
\cite{Jones07}. Among these are the classical molecular dynamics
(MD) simulation method, classical Monte Carlo integration, and
liquid state theory \cite{Hansen}, modified with effective pair
potentials that incorporate essential quantum effects such as
diffraction and degeneracy. Early approaches were based on a
classical form for the two particle density matrix in coordinate
representation to identify the effective pair potential
incorporating diffraction effects \cite{Ebeling06,Filinov04}.
Exchange effects were incorporated in a similar way using the pair
correlation function for an ideal gas \cite{Exchange,PDW}. More
recently, effective classical systems have been defined with an
effective temperature as well as pair potential \cite{PDW,Dufty12}.
A formalism for construction of a classical system with
thermodynamics and structure corresponding to a given quantum system
is described in reference \cite{DDT12}. A system of particular
interest exhibiting strong Coulomb coupling effects, both classical
and quantum, is the electron gas (referred to classically as the one
component plasma or quantum mechanically as jellium). In the
classical limit its thermodynamics is completely characterized by
the Coulomb pair potential and a dimensionless coupling constant
$\Gamma =\beta q^{2}/r_{0}$. Here $\beta =1/k_{B}T$ is the inverse
temperature, $q$ is the particle charge, and $r_{0}$ is the average
distance between particles defined in terms of the density (see
below).

The objective here is to provide an effective classical system
representing the thermodynamics of the real quantum jellium, using
an effective pair potential and an effective coupling constant.
Simple analytic expressions are given, based on the more complete
but complex results of reference \cite{DDA12}. Application of this
model is illustrated using the hypernetted chain (HNC) integral
equation of classical liquid state theory to calculate the pair
correlation function. Comparison of these calculations with the
first finite temperature restricted path integral Monte Carlo (PIMC)
simulation results reported recently \cite{Brown12} show good
agreement over a wide range of densities and temperatures.

Jellium has a broader current interest than its historical role as a
test bed for quantum many body theories. Its thermodynamic
properties, particularly the density dependence of the free energy,
provide the basis for local density approximations (LDA) in electron
density functional theory (DFT) \cite{DFT}. Fits to zero temperature
PIMC simulations have been the basis for virtually all LDA DFT for
the past thirty years. Until now there has been no corresponding
basis for an LDA at finite temperatures from either theory or
simulation. Such conditions of solid densities at temperatures
comparable to the Fermi temperature are of central interest to the
new studies of "warm, dense, matter" \cite{WDM}.

\label{sec2}The system of interest is a collection of $N$ charges
with Coulomb pair interactions $\phi(r)$ in a uniform neutralizing
background, at equilibrium \cite{Jellium}. The thermodynamic
variables are the temperature and density, $T\equiv1/k_{B}\beta$ and
$n$. A corresponding effective classical system is considered with
pair interactions $\phi_{c}(r)$ in a uniform neutralizing
background, at equilibrium with temperature and density $T_{c}$ and
$n_{c}$. The correspondence of the classical and quantum systems is
established by defining $\phi_{c}(r),T_{c},n_{c}$ as functions or
functionals of $\phi(r),T,n$ in such a way as to assure the
equivalence of selected equilibrium properties. Three such
conditions are chosen \cite{DDT12}. The first two are equivalence of
the densities and pair correlation
functions%
\begin{equation}
n_{c}=n,\hspace{0.2in}g_{c}(r,\beta_{c},n_{c}\mid\phi_{c})=g(r,\beta,n\mid
\phi).  \label{1}
\end{equation}
The remaining condition fixing $T_{c}$ is replaced by a corresponding
condition for an effective coupling constant, as discussed below.

To be useful, the condition equating pair correlation functions must be
invertible, $\phi _{c}\left( r\right) =g_{c}^{-1}(r,\beta _{c},n_{c}\mid g)$%
, which entails solution to the classical many-body problem. (this
inversion does not need to be unique; see final comments below). In
the special case of the ideal gas limit the result is known as the
Pauli potential, denoted $\phi _{c}^{(0)}\left( r\right) $. Even in
this case the inversion
cannot be accomplished exactly, but good approximations are known \cite%
{Exchange,DDA12}. The relevant dimensionless thermodynamic
parameters for the
quantum system are the temperature relative to the Fermi temperature $%
t=1/\beta \epsilon _{F}$ and $r_{s}=r_{0}/a_{B}$, the mean distance between
particles $r_{0}$ (defined by $4\pi nr_{0}^{3}/3=1$) relative to the Bohr
radius $a_{B}$. Hence the dimensionless effective potential $\phi _{c}^{\ast
}(x,t,r_{s})\equiv \beta _{c}\phi _{c}(r,\beta ,n\mid \phi )$ is written in
the form%
\begin{equation}
\phi _{c}^{\ast }(x,t,r_{s})=\phi _{c}^{(0)\ast }(x,t)+\Delta ^{\ast }\left(
x,t,r_{s}\right) ,  \label{2}
\end{equation}%
where $x\equiv r/r_{0}$. It has been recognized that the Pauli
potential depends only on $t.$

\begin{figure*}[!hbt]
\centering
    \begin{minipage}[b]{0.47\linewidth}    
    \includegraphics[width=80mm, height=70mm]{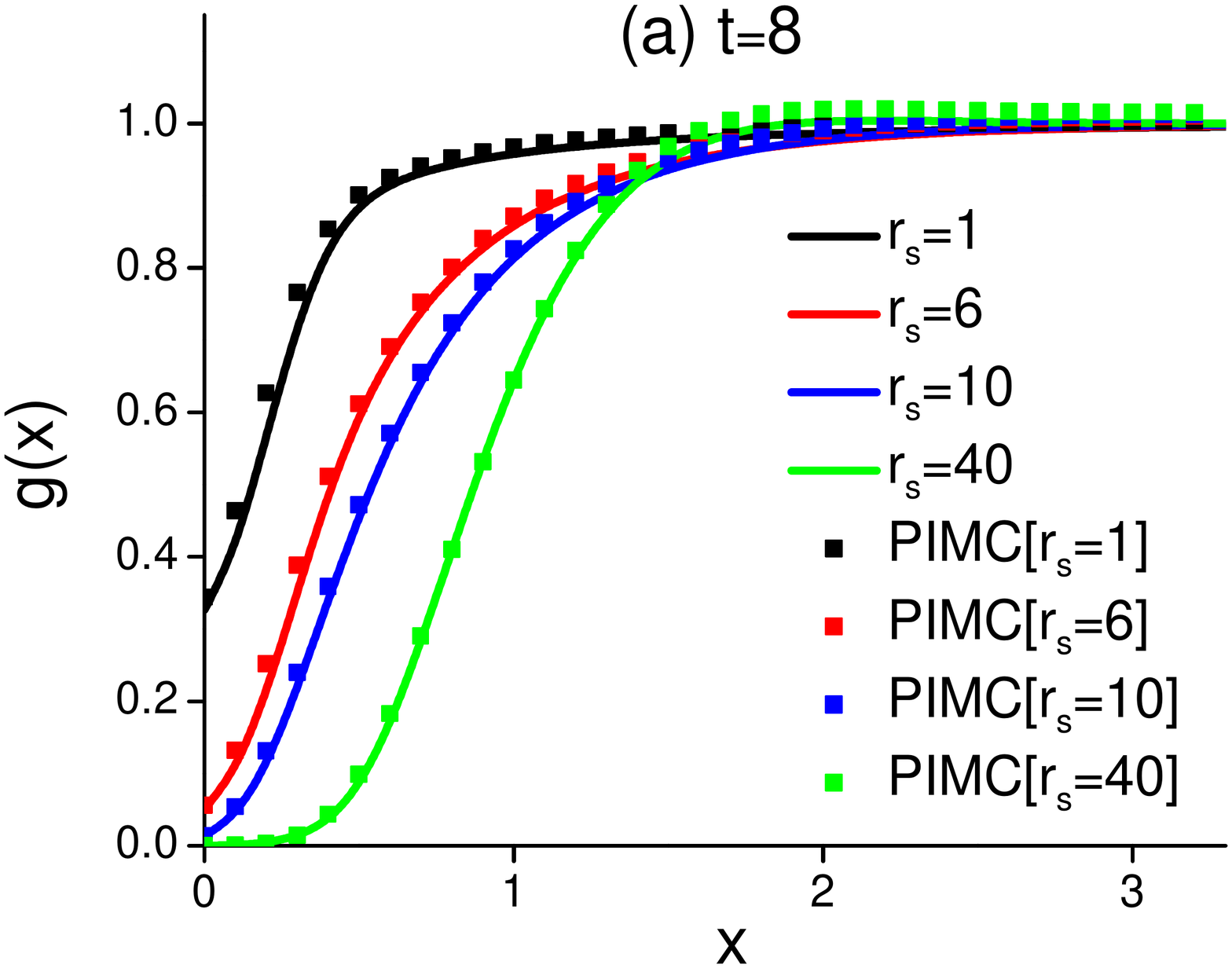}
    \label{fig1a}
    \end{minipage}
    \hspace{0.05cm}
    \begin{minipage}[b]{0.47\linewidth}
    \includegraphics[width=80mm, height=70mm]{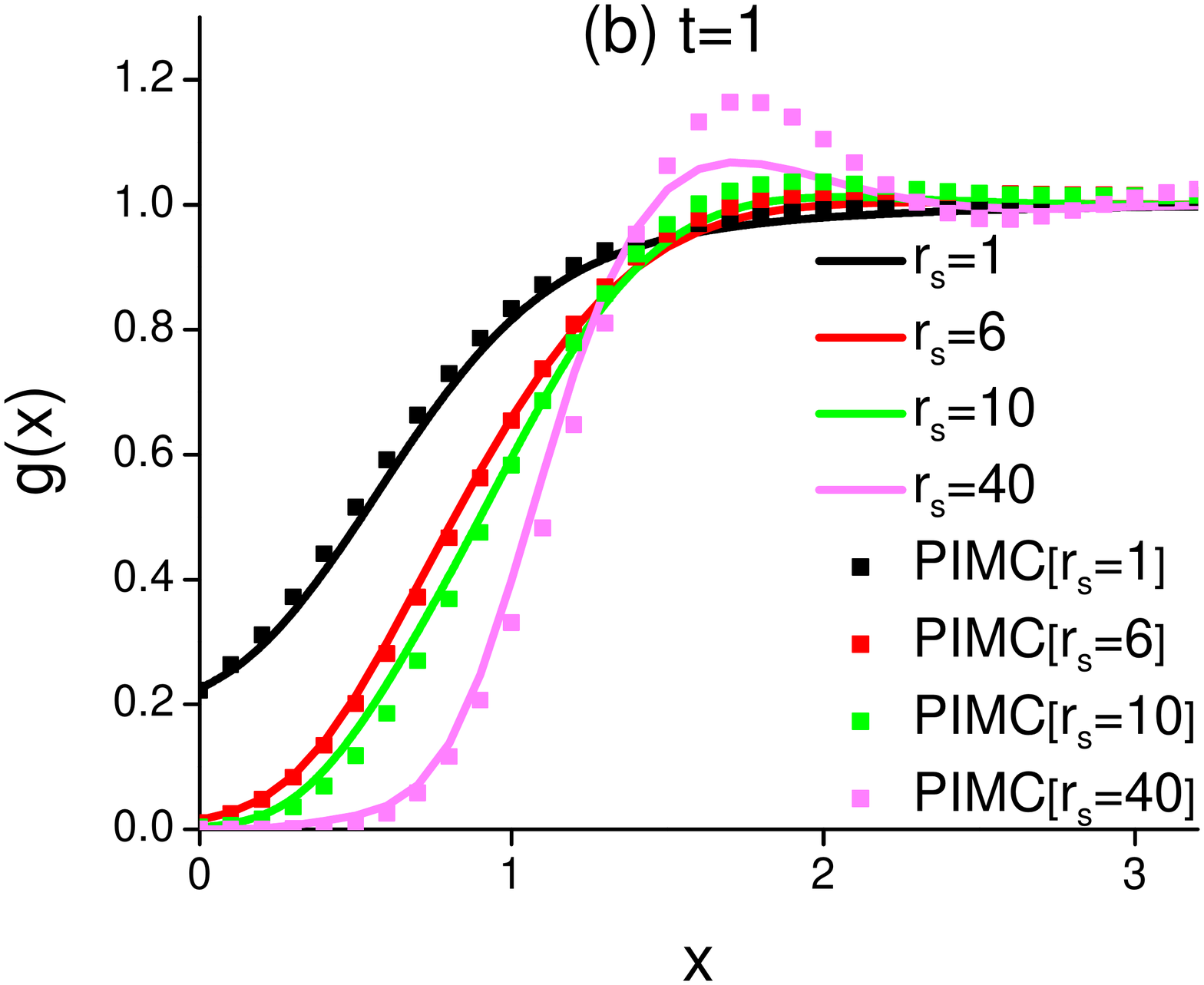}
    \label{fig1b}
    \end{minipage}
    \caption{Pair correlation function $g(r^*)$ at (a) $t = 8$ and (b) $t= 1$ for $r_s = 1, 6, 10, 40$. Also shown are the results of PIMC.}
    \end{figure*}

Two exact limits for $\Delta ^{\ast }\left( x,t,r_{s}\right) $ are important
for the discussion here. The first is the weak coupling limit
\begin{equation}
\phi _{c}^{\ast }(x,t,r_{s})\rightarrow -c_{c}(x,t,r_{s}),  \label{3}
\end{equation}%
where $c_{c}(x,t,r_{s})$ is the direct correlation function. It is
related to the pair correlation function $g_{c}(x,t,r_{s})$ by the
exact Ornstein-Zernicke equation \cite{Hansen}. Using the
correspondence conditions (\ref{1}) the Ornstein-Zernicke equation
defines the direct correlation function in terms of the quantum pair
correlation function
\begin{equation}
c_{c}(x)=g(x)-1-\frac{3}{4\pi }\int d\mathbf{x}^{\prime
}c(\left\vert \mathbf{x}-\mathbf{x}^{\prime }\right\vert )\left[
g(x^{\prime })-1\right] . \label{3a}
\end{equation}%
A sufficient condition for weak coupling is large $x$, for which the
behavior of $c_{c}(x,t,r_{s})$ is determined from the perfect screening sum
rule for $g(x,t,r_{s})$ \cite{Martin99}, giving%
\begin{equation}
\Delta ^{\ast }\left( x,t,r_{s}\right) \rightarrow \Gamma _{c}\left(
t,r_{s}\right) x^{-1}.  \label{6}
\end{equation}%
This is the same form as for the classical one component plasma,
except with the classical Coulomb coupling constant $\Gamma =\beta
q^{2}/r_{0}=$ $4\left( 2/3\pi
^{2}\right) ^{1/3}r_{s}/3t$ replaced by the effective coupling constant%
\begin{equation}
\Gamma _{c}\left( t,r_{s}\right) =\frac{2}{\beta \hbar \omega _{p}\coth
\left( \beta \hbar \omega _{p}/2\right) }\Gamma .  \label{7}
\end{equation}%
Here $\omega _{p}=\left( 4\pi nq^{2}/m\right) ^{1/2}$ is the plasma
frequency (or, equivalently, $\beta \hbar \omega _{p}=4\left( 2\sqrt{3}\pi
^{-2}\right) ^{1/3}r_{s}^{1/2}/3t$). At low temperatures and fixed density $%
\Gamma $ becomes divergent whereas the effective coupling constant remains
finite $\Gamma _{c}\left( 0,r_{s}\right) \simeq 1.\,\allowbreak
155\,r_{s}^{1/2}$. At high temperatures $\Gamma _{c}\left( t,r_{s}\right)
\rightarrow \Gamma \simeq \allowbreak 0.\,\allowbreak 543 r_{s}/t$.

The second exact limit is that for low density and weak coupling.
The condition of low density means that $g(x,,t,r_{s})$ is
determined by the two electron Slater sum. The weak coupling $\Delta
^{\ast }\left(
x,t,r_{s}\right) $ in that case is known as the Kelbg potential \cite%
{Kelbg,Filinov04,Ebeling06}
\begin{eqnarray}
\Delta ^{\ast }\left( x,t,r_{s}\right)  &\rightarrow &\Gamma x^{-1}\left(
1-\exp (-\left( ax\right) ^{2})\right.   \notag \\
&&\left. +\sqrt{\pi }(ax) erfc(ax)\right) , \label{8}
\end{eqnarray}%
where $a=\left( r_{s}/\Gamma \right) ^{1/2}$.  This weak coupling result at
low density can be improved by imposing the exact behavior of the two
particle Slater sum at $x=0$, to include some strong coupling effects \cite%
{Gombert89,Wagenknect01,Filinov04,Ebeling06}. The modified form is%
\begin{eqnarray}
\Delta _{K}^{\ast }\left( x,\Gamma ,r_{s}\right) &\equiv & \frac{\Gamma}
{x}\left(
1-\exp (\left( ax\right) ^{2})\right.   \notag \\
&&\left. +\sqrt{\pi }\frac{ax}{\gamma }erfc
(\gamma ax)\right) ,  \label{9}
\end{eqnarray}%
with%
\begin{equation}
\gamma \left( \Gamma r_{s}\right) =-\frac{\left( \pi \Gamma r_{s}\right)
^{1/2}}{\mathrm{\ln }S(\Gamma r_{s})}  \label{10}
\end{equation}%
where $S(\Gamma r_{s})$ is the two electron relative coordinate Slater sum at $x=0$%
\begin{equation}
S(\Gamma r_{s})=-4\left( \pi \Gamma r_{s}\right) ^{1/2}\int_{0}^{\infty
}dye^{-y^{2}}\frac{y}{1-e^{\pi \left( \Gamma r_{s}\right) ^{1/2}/y}}.
\label{11}
\end{equation}

The proposal here is to further extend this Kelbg form to apply broadly
across a wide range of values $t,r_{s}$ by imposing the exact asymptotic
limit (\ref{6}). This is accomplished by replacing $\Gamma $ with the
effective coupling constant  $\Gamma _{c}$ given by (\ref{7}). The
approximate effective pair potential is thus%
\begin{equation}
\phi _{c}^{\ast }(x,t,r_{s})\simeq \phi _{c}^{(0)\ast }(x,t)+\Delta
_{K}^{\ast }\left( x,\Gamma _{c},r_{s}\right) .  \label{12}
\end{equation}%
Since $\Delta _{K}^{\ast }\left( x,\Gamma _{c},r_{s}\right) $ is an
analytic, parameter free form this potential is suitable for practical
applications in classical many-body theory, classical Monte Carlo
calculations, and molecular dynamics simulations.

To illustrate the utility of this model potential the pair
correlation function $g(x,t,r_{s})$ for jellium is calculated here
using the classical liquid state HNC integral equation
\cite{Hansen}. The first step is a determination of $\phi
_{c}^{(0)\ast }(x,t)$ for the ideal Fermi gas. Since the pair
correlation function $g^{(0)}(x,t,r_{s})$ is known exactly, the HNC
equations can be inverted to determine $\phi _{c}^{(0)\ast }(x,t)$.
These equations are solved numerically using the method of Ng
\cite{Ng}. Next, with $\phi _{c}^{(0)\ast }(x,t)$ known the pair
correlation
function for jellium can be determined from the HNC equations using (\ref{12}%
).

Very recently restricted path integral Monte Carlo (PIMC)
simulations have been reported for the pair correlation function
spanning conditions ranging from extreme quantum to semi-classical.
These results provide important benchmarks for existing quantum
many-body methods, as well as the
approach proposed here. Consider first the relatively high temperature $t=8$%
. Figure (1a) shows good agreement with PIMC for all densities,
$1\leq r_{s}\leq 40$. Figure (1b) shows the same comparison for
$t=1$. Again the agreement is good, except at the extreme condition
$r_{s} = 40$. In this case a strong correlation peak has formed that
is badly under estimated by the theory, although its location is
adequately described.

Generally, it is found for $t\leq 1$ the theory is quite good for
$1\leq r_{s}\leq 10$. This is illustrated in Figures (2a) and (2b)
at $t=0.5$ and $0.0625$ (the latter is essentially the same as
$t=0$, as confirmed by a comparison with diffusion Monte Carlo
simulations at $t=0$ \cite{Ortiz}). Some trends are evident even
from this limited data. For example, the temperature dependence for
$t \leq 0.5$ is quite weak for $r_{s} > 1$. However, for $r_{s} = 1$
a significant temperature dependence is seen for $0.5 \leq t \leq
8$. In summary, the model potential (\ref{12}) provides a practical
form for the analysis of jellium using classical methods under
conditions that are difficult to access by existing quantum methods (e.g., $%
r_{s}>1$ and $t<10$).

\begin{figure*}[!ht]
\centering
    \begin{minipage}[b]{0.47\linewidth}
    \centering
    \includegraphics[width=80mm, height=70mm]{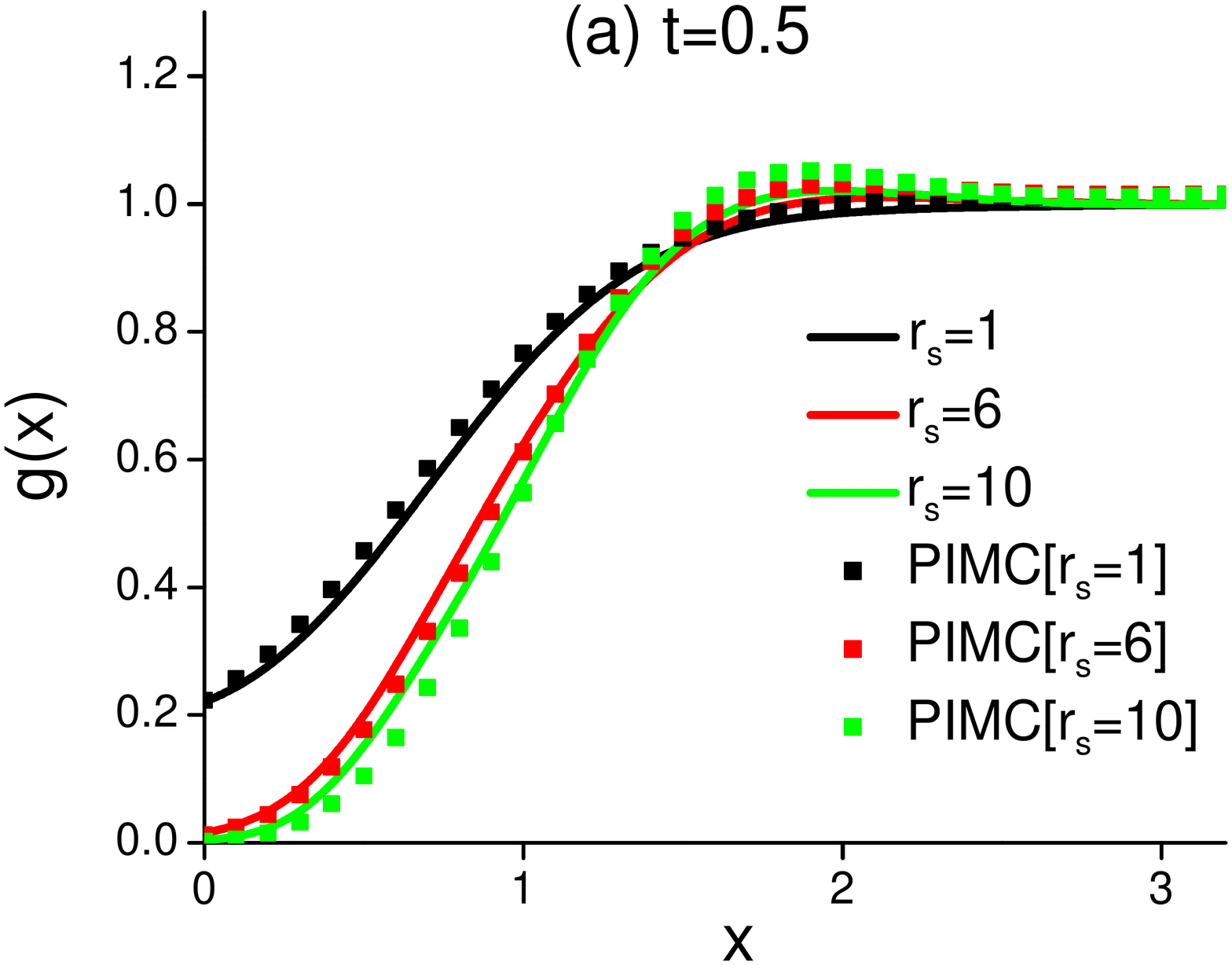}
    \label{fig2a}
    \end{minipage}
    \hspace{0.05cm}
    \begin{minipage}[b]{0.47\linewidth}
    \centering
    \includegraphics[width=80mm, height=70mm]{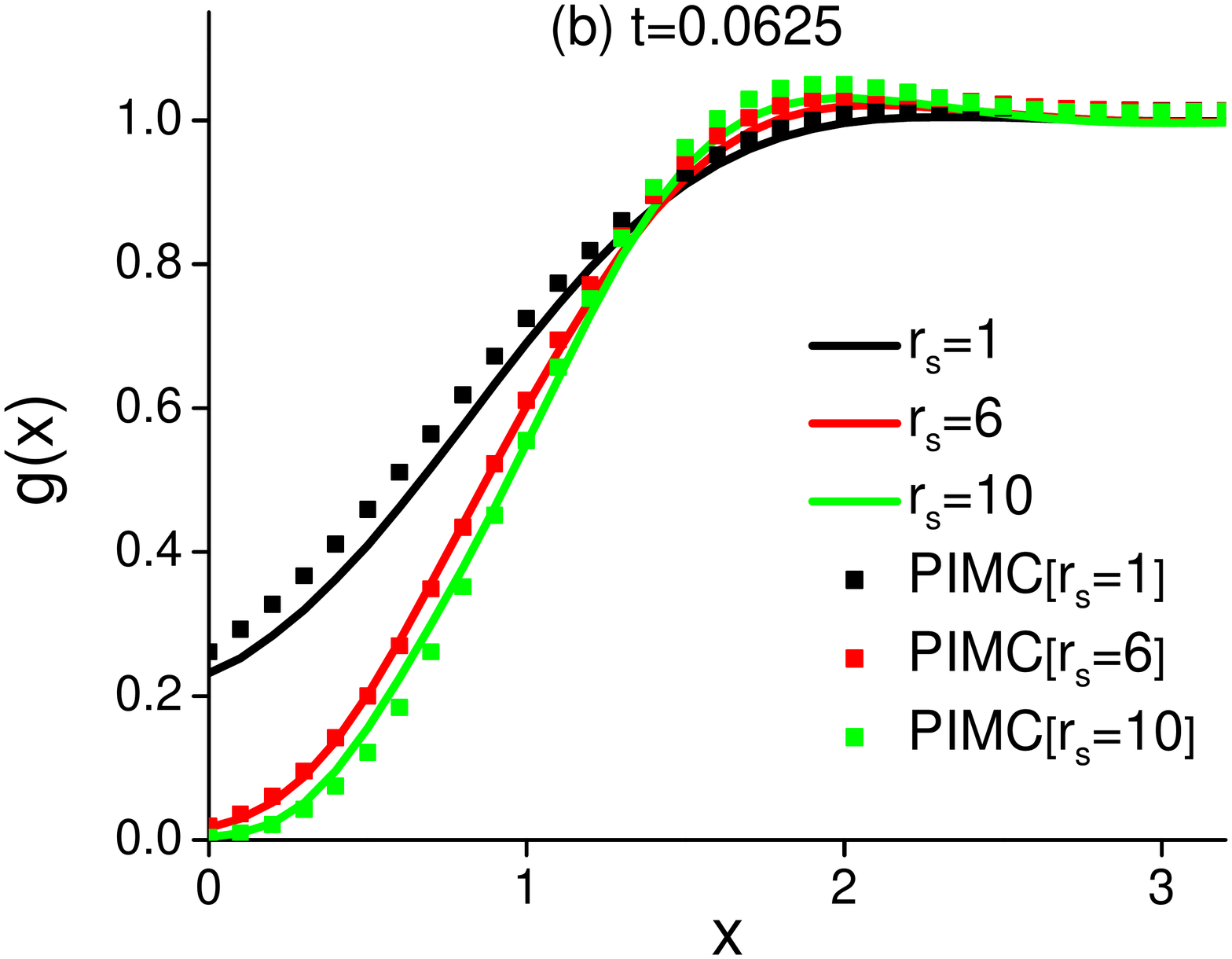}
    \label{fig2b}
    \end{minipage}
    \caption{Pair correlation function $g(r^*)$ at (a) $t=0.5$ and (b) $t=0.0625$ for $r_s = 1, 6, 10$. Also shown are the results of PIMC.}
    \end{figure*}

The thermodynamic properties of jellium can be calculated from the
pair correlation function. For example, the pressure can be obtained
from a coupling constant integration. Let $p(t,r_{s},q)$ be the
exact quantum pressure and $g\left( r,t,r_{s},q\right) $ the exact
quantum pair correlation function where now the dependence on the
charge $q$ has been made explicit. Then
\begin{eqnarray}
p(t,r_{s},q) &=&p(t,r_{s},0)+8\pi \int_{0}^{q}dyy  \notag \\
&&\times \int_{0}^{\infty }drr^{2}\phi (r)\left( g\left(
r,t,r_{s},y\right) -1\right) .  \label{13}
\end{eqnarray}%
Here $\phi (r)$ is the Coulomb pair potential. Therefore, approximating $%
g\left( r,t,r_{s},y\right) $ by the corresponding classical result obtained
using the model potential (\ref{12}) determines the pressure for arbitrary $%
t,r_{s}$. A more direct approach would be classical Monte Carlo
integration of the Gibbs distribution for the free energy
\begin{equation}
F=-\beta ^{-1}\ln r_{0}^{N}\int d\mathbf{x}_{1}..d\mathbf{x}%
_{N}e^{-\sum\limits_{ij}\left( \phi _{c}^{(0)\ast }(x_{ij},t)+\Delta
_{K}^{\ast }\left( x_{ij},\Gamma _{c},r_{s}\right) \right) },
\label{13a}
\end{equation}%
with $x_{ij}=\left\vert \mathbf{x}_{i}-\mathbf{x}_{j}\right\vert .$

As noted in the introduction, the idea of an effective classical
pair potential with an effective classical temperature was already
introduced more than ten years ago by Perrot and Dharma-wardana
\cite{PDW}. Instead of the\ Kelbg potential they chose the Deutsch
potential \cite{Deutsch}, originally introduced as a simpler
representation of the Kelbg potential.
The PDW effective classical potential is similar to (\ref{12}), but with $%
\Delta _{K}^{\ast }\left( x,\Gamma _{c},r_{s}\right) $ replaced by %
\begin{equation}
\Delta _{PDW}^{\ast }\left( x,\Gamma _{PDW},r_{s}\right) \equiv
\Gamma _{PDW}x^{-1}\left( 1-\exp (-bx)\right)   \label{14}
\end{equation}%
Here, $b=\left( \pi r_{s}/\Gamma _{PDW}\right) ^{1/2}$ and the effective
coupling constant is%
\begin{equation}
\Gamma _{PDW}=\left( 1+\left( \frac{T_{0}}{T}\right) ^{2}\right)
^{-1/2}\Gamma   \label{15}
\end{equation}%
This follows from their phenomenological form for the classical
temperature interpolating between the real temperature T and a
finite value $T_{0}$ at $T=0$. The single parameter $T_{0}/T$ is
determined by fitting the classical correlation energy calculated
with this potential to the quantum exchange/correlation energy
determined from PIMC at T=0. It is given as an explicit fitting
function of $r_{s}$ in reference \cite{PDW}. Although the dependence
of $\Gamma_{PDW}$ on $t,r_{s}$ is quite different from that derived
here, and the shape of the resulting effective pair potential can be
quite different, nevertheless the HNC pair correlation function
calculated from the PDW potential has a similar accuracy to that
reported here. This indicates that an effective pair potential has
no inherent physical interpretation, but rather is a non-unique tool
for generating physical properties of interest through classical
many-body methods. Here that potential has been constructed by
imposing three exact constraints: the ideal gas limit, low density
limit, and large distance limit. Consequently no fitting parameters
are required. The result provides theoretical support for the ideas
of reference \cite{PDW} and provides insight into the relevant
physical mechanisms. For example, the exact screening sum rule that
determines the form of $\Gamma _{c}$ here appears to incorporate
quantum effects as significant as those of $\Gamma _{PDW}$ imposed
by empirical $T=0$ exchange/correlation energy data.

This research has been supported by NSF/DOE Partnership in Basic Plasma
Science and Engineering award DE-FG02-07ER54946 and by US DOE Grant
DE-SC0002139.


\bigskip

\begin{thebibliography}{99}
\bibitem{Jones07} C. Jones and M. Murillo, High Energy Density Physics
\textbf{3}, 379 (2007); F. Graziani et al, High Energy Density Physics
\textbf{8},105 (2012).

\bibitem{Hansen} J-P Hansen and I. MacDonald, \emph{Theory of Simple Liquids}%
, (Academic Press, San Diego, CA, 1990).

\bibitem{Ebeling06} For references see W. Ebeling, A. Filinov, M. Bonitz, V.
Filinov, and T. Pohl, J. Phys. A \textbf{39}, 4309 (2006).

\bibitem{Filinov04} A. Filinov, V. Golubnychiy, M. Bonitz, W. Ebeling, and
J. Dufty, Phys. Rev. E \textbf{70,} 046411 (2004).

\bibitem{Exchange} G.E. Uhlenbeck, L. Gropper, Phys. Rev. \textbf{41} (1932)
79; F. Lado, J. Chem. Phys. \textbf{47}, 5369 (1967); J. W. Dufty, S. Dutta,
M. Bonitz, and A. Filinov, Int. J. Quant. Chem. \textbf{109}, 3082 (2009).

\bibitem{PDW} F. Perrot and M. W. C. Dharma-wardana, Phys. Rev. B \textbf{62}%
, 16536 (2000); M. W. C. Dharma-wardana, Int. J. Quant. Chem. \textbf{112},
\textbf{53} (2012).

\bibitem{Dufty12} J. W. Dufty and S. Dutta, Contrib. Plasma Phys. \textbf{52}%
, 100 (2012).

\bibitem{DDT12} J. W. Dufty and S. Dutta, "Classical Representation of a
Quantum System at Equilibrium: Theory", Phys. Rev. E (to appear).

\bibitem{DDA12} S. Dutta and J. W. Dufty, "Classical Representation of a
Quantum System at Equilibrium: Application", Phys. Rev. E (to
appear).

\bibitem{Brown12} E. Brown, B. Clark, J. DuBois, D. Ceperley, \emph{Path
Integral Monte Carlo Simulation of the Warm-Dense Homogeneous Electron Gas }%
, cond mat arXiv:1211.6130, 2012.

\bibitem {DFT} \emph{Density Functional Theory: An Advanced Course}, E. Engel and R.M. Dreizler (Springer, Heidelberg, 2011).

\bibitem{WDM} V.V. Karasiev, T. Sjostrom, D.Chakraborty, J.W. Dufty, F.E.
Harris, K. Runge, and S.B. Trickey, "Innovations in
Finite-Temperature Density Functionals", chapter in
\emph{Computational Challenges in Warm Dense Matter}, F. Graziani et
al. eds., Springer Verlag, (to appear); \emph{Basic Research Needs
for High Energy Density Laboratory Physics} (Report of the Workshop
on Research Needs, Nov. 2009). U.S. Department of Energy, Office of
Science and National Nuclear Security Administration (2010); see
Chapter 6 and references therein.

\bibitem{Jellium} D. Kremp, M. Schlanges, W. Kraeft, \emph{Quantum
Statistics of Nonideal Plasmas}, (Springer-Verlag, Berlin, 2005); G.
Giuliani and G. Vignale, \emph{Quantum Theory of the Electron Liquid},
(Cambridge U. Press, Cambridge, 2005).

\bibitem{Kelbg} G. Kelbg, Ann. Phys. \textbf{12}, 219 (1963).

\bibitem{Gombert89} M.-M. Gombert and H. Minoo, Contrib. Plasma Phys.
\textbf{29}, 355 (1989).

\bibitem{Wagenknect01} H. Wagenknecht, W. Ebeling, and A. F\"{o}rster,
Contrib. Plasma Phys. \textbf{41}, 15 (2001).

\bibitem{Deutsch} C. Deutsch, Phys. Lett. A \textbf{60,} 317 (1977); H.
Minoo, M. Gombert, and C. Deutsch, Phys. Rev. A \textbf{23}, 924 (1981).

\bibitem {Martin99}D. Pines and Ph. Nozieres, \emph{The Theory of Quantum Liquids}, (Benjamin, NY, 1966);
D. Brydges and Ph. Martin, J. Stat. Phys. \textbf{96,} 1163 (1999).

\bibitem{Ng} K-C Ng, J. Chem. Phys. \textbf{61}, 2680 (1974).

\bibitem{Ortiz} G. Ortiz and P. Ballone, Phys. Rev. B \textbf{50}, 1391
(1994).

\end{thebibliography}
\end{document}